# Towards Generating Benchmark Datasets for Worm Infection Studies

Sara Asgari
*Department of Computer Engineering*
*Amirkabir University of Technology*
Tehran, Iran
s.asgari@aut.ac.ir

Babak Sadeghiyan
*Department of Computer Engineering*
*Amirkabir University of Technology*
Tehran, Iran
basadegh@aut.ac.ir

*Abstract*— Worm origin identification and propagation path reconstruction are among the essential problems in digital forensics. Until now, several methods have been proposed for this purpose. However, evaluating these methods is a big challenge because there are no suitable datasets containing both normal background traffic and worm traffic to evaluate these methods. In this paper, we investigate different methods of generating such datasets and suggest a technique for this purpose. ReaSE is a tool for the creation of realistic simulation environments. However, it needs some modifications to be suitable for generating the datasets. So we make required modifications to it. Then, we generate several datasets for Slammer, Code Red I, Code Red II and modified versions of these worms in different scenarios using our technique and make them publicly available.

## I. INTRODUCTION

Worm origin identification and propagation path reconstruction are among the essential problems in digital forensics and helps us to understand the cause of risks and do better security measures. However, the evaluation of these methods is problematic. Both normal background traffic and worm traffic is required to evaluate these methods. Currently, in the research community to assess these methods, worm traffic is injected into normal background traffic that is generated separately, e.g. in [1] and [2] this approach is used to generate datasets to evaluate their proposed method. In this approach, the competition of worm propagation traffic with normal background traffic over bandwidth is not considered because these two traffic is not generated in the same network and simultaneously. Besides, obtaining normal background traffic is one of the main problems in evaluating worm origin identification methods because network administrators often avoid sharing their network traffic because of privacy concerns. Besides, because of privacy concerns, existing datasets are usually anonymized, e.g. MAWI[3], SANTA[4], LBNL[5], PUF[6], UNIBS[7], UGR 16[8], Kent[9][10]. Currently, there are no datasets containing both normal background traffic and worm traffic that both traffic is generated simultaneously and in the same network, and those datasets can be used to evaluate worm origin identification methods. If such datasets existed, the competition of worm propagation traffic with normal background traffic over bandwidth would be considered. Also, the main challenge of researchers for evaluating worm origin identification methods that is obtaining normal background traffic would be solved.

The purpose of this paper is to suggest a technique to generate such datasets and to generate a number of such datasets. To do this, we examine different methods of generating datasets containing normal background traffic and worm traffic. For obtaining normal background traffic, we analyze 21 datasets and different traffic generators. We also study different technologies of creating worm experimental environment. The result of our investigations is that ReaSE[11] is a suitable tool for generating the datasets mentioned above. However, it needs some modifications. So we made required modifications in ReaSE and then we generate several datasets for Code Red I and Code Red II[12], Slammer[13] and modified versions of them in different scenarios.

The rest of this paper is organized as follows: Section II presents a discussion on possible methods of generating datasets for evaluating worm origin identification and propagation path reconstruction methods. Section III presents the proposed technique. This section briefly overviews ReaSE and its capabilities and the modifications we made to ReaSE. Section IV describes our generated datasets. Section V concludes the paper.

## II. DISCUSSION ON GENERATING DATASETS FOR WORM INFECTION STUDIES

To evaluate worm origin identification and propagation path reconstruction methods, we need datasets containing both normal background traffic and worm traffic. There are two approaches to generate such datasets:

1) Combining worm traffic and normal background traffic that is generated separately.
2) Generating worm traffic and normal background traffic in the same network and simultaneously.

Also, our observations indicate that there are three ways to obtain normal background traffic and worm traffic: using existing datasets, obtaining traffic of real networks, generating traffic in the experimental environment. In this section, we investigate these three ways.

### B. Using Existing Datasets

In this paper, we study 21 datasets to obtain normal background traffic. The results are as below:

- Some of these datasets, e.g. DARPA[14][15][16], KDD Cup[17], NSL-KDD[18] and PU-IDS[19] are outdated, and they are not suitable for today's evaluations because of the many modifications that have occurred in the traffic of networks so far.

- Some of them, e.g. UGR 16, SSENET-2011[20], SSENET-2014[21], SANTA, UNIBS, Unified Host and Network[22], PUF, PU-IDS, NSL-KDD, Kent and KDD Cup, are not available in the packet-based format. However, to evaluate worm origin identification and propagation path reconstruction methods, we usually need network traffic in packet-based format.

- Some datasets, e.g. SANTA and IRSC[23], are not publicly available due to privacy concerns.

- In some of them, e.g. MAWI, TUIDS[24][25], UNSW-NB15[26], NGIDS-DS[27] and CTU-13[28], normal traffic and attack traffic are mixed.

- In some datasets, e.g. LBNL, some required information (such as network topology, link delay, and bandwidth) to inject worm traffic to normal traffic is not available.

- Some of them don't include some new network protocols. For example, ISCX[29] doesn't include HTTPS.

- The testbed of some datasets, e.g. CICIDS[30] and ISCX, is too small.

- In most datasets, some packet information such as IP (e.g.in Unified Host and Network, PUF, UNIBS, MAWI, UGR 16, Kent, LBNL), payload (e.g. in CTU-13, MAWI, SANTA, LBNL), time (e.g. in Unified Host and Network, Kent) and ports (in Kent) are anonymized because of potential security risks and privacy concerns. In most datasets, the IP address is anonymized by preserving the prefix.

## C. Obtaining Traffic of Real Networks

Network administrators usually refuse to share their network traffic due to privacy concerns, and if they do so, they will usually anonymize the traffic.

## D. Generating Traffic in Experimental Environment

1) *Normal Background Traffic:* Most traffic generators generate traffic in one direction. Some others, although generate traffic in both directions but both sides of communication are not synchronous. We need two-directional traffic that each side of communication affects the other side (for example, retransmitting packets in TCP, request-response communications in application layer protocols, etc.). Therefore, although the traffic generated by many traffic generators is suitable for performance evaluation, optimization, measurement of available bandwidth and quality of service, etc., it is not useful for evaluating worm origin identification and propagation path reconstruction methods. We discuss some traffic generators in the following:

- Swing[31][32]: Swing captures the packet interactions of applications and extracts distributions for application, user, and network behaviour. Then, according to the underlying model, it generates traffic in an emulated environment. One of the limitations of swing is that it generates realistic traces for a single link.

- Tmix: Weigle et al.[33] proposed tmix for NS-2 simulator[34]. Tmix extracts the source-level characterization of TCP connections from a packet header trace that is obtained from an arbitrary link and then emulate the socket-level behaviour of the source applications that created the connections in the trace. Adurthi et al.[35] also implemented tmix for GTNetS[36]. Like swing, Tmix generates traffic for a single link.

- Packmime-HTTP: Cao et al.[37] proposed a model for HTTP traffic generation and implemented it for NS-2 simulator and called it Packmime-HTTP. Cheng et al.[38] extended and modified Packmime-HTTP in NS-2 to work with NS-3[39]. They also added an extra working mode to that. These traffic generators generate only HTTP traffic.

- Ammar et al.[40] implemented a tool to generate traffic based on Poisson Pareto Burst Process (PPBP) model [41] for NS-3 simulator. This traffic generator generates traffic only in one direction and doesn't include request-response interactions in application layer protocols.

- NeSSi [42]: NeSSi is a network simulation tool that includes different features related to network security. NeSSi is built upon the JIAC[43] framework and has provided a distributed and scalable architecture. NeSSi contains a limited number of application layer protocols.

2) *Worm Traffic:* In [44], the technologies of creating worm experimental environment are divided into several categories: analytical model, packet-level simulation, network emulation, hardware testbed, and hybrid method. As discussed in [44], The fidelity of analytical models is insufficient to use for our purpose. Using network emulation doesn't satisfy the scalability requirement of worm researches. Providing hardware testbed is not feasible because worm experiments require a large number of nodes. Packet-level simulation provides good scalability as well as better fidelity than the analytical model. Some network simulators that enable packet-level worm simulation are discussed briefly below:

- SSFNET: Liljenstam et al.[45] extended the SSFNET simulator [46][47] to include behavioural models of internet worms. They use the detailed packet-level simulation for part of the network, and a model that is less accurate but computationally efficient for other parts[48].

- NS-2: NS-2 includes some behavioural models of worms. The approach used by NS-2 is similar to that used by SSFNet (mentioned above). Also, in NS-2, there is no mechanism to assign IP addresses to nodes, and so the scanning worm models are problematic[48].

- GTNetS: In GTNetS, we can simulate different worms by setting parameters such as transport layer

protocols, infection length, scan rate, number of simultaneous connections, and so on.

- PAWS [49]: PAWS is a distributed worm spread simulator. It simplified worm propagation behaviour to enhance simulation speed. For example, only those scans are delivered to the destination that the destination host has not yet infected. Also, to reduce the overhead, PAWS aggregates the packets that want to be sent to a node in a single message and send the message at the end of the time unit.

- Wagner et al. [50] implemented a simulator for worm propagation in the Perl scripting language. Although this simulator considers the effect of link bandwidth and propagation delays, but ignores queuing, loss and competing traffic. Also, the TCP model is used by TCP-based worms is simplified that doesn't consider some features such as slow start and congestion window [48].

- NeSSi: In addition to normal background traffic generation, Nessi also makes it possible to simulate worm propagation. The worm propagation scheme provided by Nessi enables the simulation of SQL Slammer and Blaster worms. Additionally, the researchers can extend the worm model provided by NeSSi or develop a new one.

III. OUR PROPOSED TECHNIQUE

The result of our investigations is that the suitable tool for generating our datasets is ReaSE. However, it needs some modifications. In this section, we first overview ReaSE and its capabilities then we express the modifications we made to it.

*A. ReaSE Overview*

OMNeT++[51] is a modular, component-based C++ simulation library and framework, primarily for building network simulators.

INET[52] framework is an extension of OMNeT++ that supports simulations of common internet protocols (such as UDP, TCP, IP, and ICMP) and intermediate and end systems (such as routers and hosts) and is used to simulate internet-like networks.

ReaSE is a tool developed to create a realistic simulation environment in combination with INET framework. It considers three aspects:

1) *Topology generation:* ReaSE uses positive-feedback preference model (PFP)[53] to generate realistic topologies. The PFP model randomly generates topologies that show power-law distribution in node degrees. In ReaSE, topology generation is divided into two parts. First, the connections of the ASs (Autonomous Systems) are created. Then the topology inside each AS is generated. The hierarchical structure generated for each AS consists of three layers, including core, gateway, and edge.

2) *Normal traffic generation:* In ReaSE, realistic normal background traffic is generated between hosts. Realistic in this case means that the generated traffic exhibits self-similar behaviour[54] and is based on a combination of different types of traffic. In ReaSE, eight traffic profiles (backup, interactive, web, mail, nameserver, streaming, Misc, and ping) based on different transport layer protocols are defined. The behaviour of each traffic profile is determined by setting various parameters (such as Request Length, Reply Length, Requests Per Flow, Time Between Requests, Reply Per Request, Time To Respond, Time Between Flows, Selection Probability, WAN Probability). So we can generate different normal background traffic patterns by using different traffic profiles and by setting different values for the traffic profile parameters.

3) *Attack traffic generation:* ReaSE makes it possible to simulate some network attacks, including worm propagation. Although it has been stated in [11] that both UDP-based and TCP-based worms have been implemented in ReaSE, TCP-based worm is not publicly available. However, by setting parameters such as payload length, time between probing packets, infection port, the range of IP addresses to scan, we can simulate different UDP worms.

*B. ReaSE Modifications*

Below, we briefly explain the modifications we made to ReaSE and don't discuss technical details:

- In ReaSE, there are generally two types of nodes: nodes that generate normal background traffic and nodes that generate worm traffic. In other words, nodes that generate normal background traffic are not capable of generating worm traffic and vice versa. But we need to generate worm traffic and normal background traffic simultaneously and by the same nodes to take into account the competition of normal traffic and worm traffic over bandwidth. So we add the module that implements worm propagation functionality (e.g. udpWormVictim.ned in UDP-based worms) in the nodes that generate normal traffic and remove the nodes that only generate worm traffic from the simulator.

- In real networks, a server usually plays two roles:
  - listening on a specific port and responding to requests
  - sending requests to other servers if needed (For example, a DNS server often sends requests to other DNS servers in the network to respond to client requests)

  In ReaSE, servers only play the first role. We change them to play both roles. To do this, we add InetUser module of ReaSE to the servers and make some necessary modifications to InetUser and ConnectionManager modules and InetUser class. So each server can also communicate with other servers.

- We use the TCPDump module of INET and provide the ability to capture the traffic of each node in pcap format.

- In ReaSE, a source port is randomly selected for each traffic profile, and the specified source ports of traffic profiles are used for generating traffic. However, in real networks, hosts use a random source port each time they establish a connection. So we change TrafficProfile struct of ReaSE to consider this.

- In ReaSE, the port on which each server listens to requests is not necessarily unique. We set the port parameter of each server so that each of them listens on a unique port.

- Due to network traffic modifications, we add four traffic profiles, including HTTP, HTTPS, SSH, and FTP, to eight traffic profiles of ReaSE that can be used to generate normal background traffic. We also create a server for each of these traffic profiles (except SSH). To do this, we implement a compound module for each server (i.e. HTTPServer.ned, HTTPSServer.ned, and FTPServer.ned).

- As previously mentioned, the TCP-based worm implemented in ReaSE is not publicly available. TCP-based worms establish a TCP connection to each target machine before sending payload packets. Besides, they use several simultaneous TCP connections to infect several machines simultaneously and speed up the propagation process. So by considering behaviours of TCP-based worms, we implement a model that different TCP-based worms that their scanning techniques are local preference scanning or uniform random scanning and their epidemiological model are SI (Susceptible-Infected) or SIR (Susceptible-Infected-Recovered) can be simulated by setting parameters such as the number of simultaneous connections, infection length, infection port, recovery probability, preference probability, and IP address range for probing packets. We implement two classes named TCPWorm and TCPWormThread. TCPWorm class stores threads in a list and manages them. TCPWormThread class handles each individual thread. We also implement a module, TCPWorm, that consists of these two classes. We add this module to vulnerable hosts.

- The scanning technique of the UDP-based worm model implemented in ReaSE is uniform random scanning. We change this model such that the scanning technique can be either local preference scanning or uniform random scanning, and the preference probability can be set. Also, this model is based on SI, so we change the epidemiological model to be SI or SIR, and the recovery probability can be set. In ReaSE, the infected host always uses a fixed source port to send probing packets. We change the UDP worm model to use random source ports for scanning IP addresses.

*C. Generating Datasets*

Using the technique described in this section, different datasets containing both normal background traffic and worm traffic can be generated for evaluating worm origin identification and propagation path reconstruction methods. Different traffic patterns of normal traffic can be generated using different traffic profiles and by setting different values for traffic profile parameters. Besides, different types of worms that their scanning strategy is preferential scanning or uniform random scanning can be simulated just by setting a number of parameters. However, other types of scanning worms, e.g. sequential scanning worms, can be simulated by making simple modifications in ReaSE. Also, different topologies can be generated by setting different parameters.

*D. Validation*

In [55], several experiments have been performed to validate that the topologies and background traffic generated by ReaSE show realistic characteristics. The results of these experiments are as below:

- the topologies generated by ReaSE show the power-law distribution in node degree well, although there are some deviations in a small number of nodes.

- The generated background traffic shows self-similarity.

Note that we didn't make any modifications to the topology generation capability of ReaSE. In addition, ReaSE uses two mechanisms to achieve self-similar traffic behaviour:

1) Using several traffic sources that are switched on and off based on heavy-tailed intervals.

2) Using heavy-tailed packet sizes for different traffic flows.

We didn't interfere in the operation of these mechanisms. So the self-similarity of normal background traffic has been preserved.

IV. OUR GENERATED DATASETS

We generate two categories of datasets that each of them contains several sets of traffic traces. In this section, we explain how we generate these datasets. We will make these datasets publicly available as soon as possible on *https://github.com/Sara-Asgari/Datasets*.

There are two approaches to generate realistic topologies[56]:

1) generating topology based on observations of real networks

2) random topology generation

The topology generated using the first approach is very realistic. Also, Random topology generation is widely used in the research community[11]. We use both approaches to generate our datasets.

*A. Category I*

Our created topology has been derived from a simplified version of the topology of a large ISP in Italy and its bandwidths and link lengths that is stated in [57], network properties used in [58] and by considering current network topologies. Fig. 1. shows the proposed topology, and Table I presents bandwidths and delays of links in the proposed topology. This network is composed of 4 core nodes, 8 gateway nodes, 16 edge nodes, and 200 end-hosts that are located in four subnets. As illustrated in Fig. 1., in this network, some redundancies have been considered to increase availability.

To generate normal background traffic, we extract the type and percentage of application layer protocols of the first-day traffic of the CICIDS dataset (illustrated in Table II) and use these values to choose the traffic profiles and the selection probability of each one. We also extracted the approximate values of the parameters (on average) for each of these traffic profiles from this dataset.

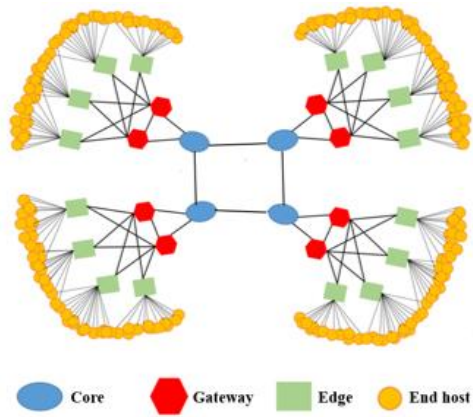

Fig. 1. Proposed topology in category I.

TABLE I. LINK PARAMETERS IN THE PROPOSED TOPOLOGY IN CATEGORY I

| Link | Bandwidth | Delay |
|---|---|---|
| Core to Core | 50 Gbps | 3 ms |
| Core to Gateway | 20 Gbps | 2 ms |
| Gateway to Edge | 10 Gbps | 0.25 ms |
| Edge to Server | 2.5 Gbps | 5 µs |
| Edge to Client | 100 Mbps | 5 µs |

TABLE II. TRAFFIC PROFILES WITH PERCENTAGES IN CATEGORY I

| Traffic Profile | % |
|---|---|
| HTTP | 53.85% |
| HTTPS | 38.13% |
| DNS | 6.87% |
| SSH | 0.78% |
| FTP | 0.20% |
| Email | 0.14% |
| Ping | 0.03% |

In category I, we generate six sets of traffic traces. The infection network and worm propagation parameters in these sets are presented in Table V. In each set, the specified worm propagates three times with different origins and propagation paths in the network. The simulated worms are Code Red I, Code Red II, quasi Code Red II, Slammer and quasi Slammer. The scanning technique of quasi Slammer is local preference while Slammer worm use uniform random scanning technique. Also, the preference probability parameter of quasi Code Red II is different from the original version of Code Red II. In our simulations, we focus on spreading part of worms and ignore their attacking part.

In simulated TCP worms, the number of concurrent connections is much lower than the value of this parameter in the original Code Red I and Code Red II because the original versions of Code Red have been propagated in Internet containing several thousands of machines, but the number of nodes in our simulations is much lower. So setting large values for this parameter causes the network to become infected in a fraction of a second that is usually not suitable for evaluations. Also, in our simulated UDP worms, the time between probing packets is much greater than the value of this parameter in the original version of Slammer worm.

*B. Category II*

In category II, we use topology generation capability provided by ReaSE for creating the network topology. This topology is shown in Fig. 2. also, Link parameters in this topology are presented in Table III. This network is composed of 10 core nodes, 20 gateway nodes, 152 edge nodes, and 1162 end-hosts and consists of 10 subnets.

The type and percentage of application layer protocols used in normal traffic generation are shown in Table IV. Also, the values set for traffic profile parameters are different from category I.

In category II, we generate two sets of traffic traces. The simulated worms are the same as category I that some parameters of them are changed (illustrated in Table V).

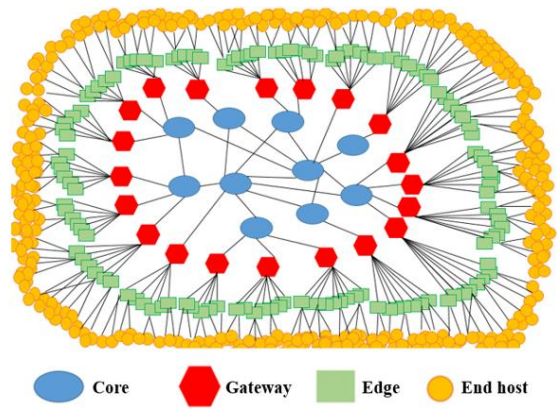

Fig. 2. Proposed topology in category II.

TABLE III. LINK PARAMETERS IN THE PROPOSED TOPOLOGY IN CATEGORY II

| Link | Bandwidth | Delay |
|---|---|---|
| Core to Core | 40 Gbps | 4 ms |
| Core to Gateway | 16 Gbps | 2.5 ms |
| Gateway to Edge | 8 Gbps | 0.3 ms |
| Edge to Server | 2 Gbps | 10 µs |
| Edge to Client | 80 Mbps | 10 µs |

TABLE IV. TRAFFIC PROFILES WITH PERCENTAGES IN CATEGORY II

| Traffic Profile | % |
|---|---|
| HTTPS | 49.2% |
| HTTP | 35.5% |
| DNS | 8.9% |
| FTP | 3.3% |
| Email | 2.8% |

TABLE V. OUR GENERATED DATASETS

| | dataset | Worm parameters | | | | | | | Infection network parameters | |
|---|---|---|---|---|---|---|---|---|---|---|
| | | *Name* | *Transport layer protocol* | *Number of concurrent connection (TCP worms)* | *Time between probing packets (UDP worms)* | *Scanning strategy* | *Preference probability (local preference scanning worms)* | *Recovery Probability* | *Number of nodes* | *Type of nodes* |
| Category I | Set 1 | Slammer | UDP | - | Uniform(4ms, 8ms) | Uniform Random | - | $10^{-4}$ per ms | 30 | HTTP Server, HTTPS Server and Client |
| | Set 2 | Quasi Slammer | UDP | - | Uniform(5ms, 10ms) | Local Preference | $\frac{1}{8}$: random $\frac{4}{8}$: same class A $\frac{3}{8}$: same class B | $10^{-4}$ per ms | 28 | HTTP Server |
| | Set 3 | Quasi Slammer | UDP | - | Uniform(5ms, 10ms) | Local Preference | 0.3: random 0.7: same subnet | $10^{-4}$ per ms | 35 | Client |
| | Set 4 | Code Red I | TCP | 23 | - | Uniform Random | - | $10^{-4}$ per ms | 28 | HTTP Server |
| | Set 5 | Code Red II | TCP | 25 | - | Local Preference | $\frac{1}{8}$: random $\frac{4}{8}$: same class A $\frac{3}{8}$: same class B | $10^{-4}$ per ms | 28 | HTTP Server |
| | Set 6 | Quasi Code Red II | TCP | 25 | - | Local Preference | 0.3: random 0.7: same subnet | $10^{-4}$ per ms | 35 | Client |
| Category II | Set 1 | Quasi Code Red II | TCP | 20 | - | Local Preference | 0.3: random 0.7: same subnet | $10^{-5}$ per ms | 52 | HTTP Server |
| | Set 2 | Quasi Slammer | UDP | - | Uniform(10ms, 12ms) | Local Preference | 0.3: random 0.7: same subnet | $10^{-5}$ per ms | 52 | HTTP Server |

## V. CONCLUSION

In this paper, we suggested a technique for generating datasets containing both normal background traffic and worm traffic that can be used to evaluate worm origin identification and propagation path reconstruction methods. To do this, we investigated different methods of generating such datasets. The result of our investigations was that ReaSE is a suitable tool for this purpose but needs some modifications. Using this technique, different datasets (different types of worms, topologies and normal background traffic) can be generated. In addition, we discussed the validation of the generated topologies and normal traffic and explained that they show realistic characteristics. We also generated multiple datasets for both UDP-based and TCP-based worms with different parameters in two different networks with different background traffic.